\documentstyle[12pt,epsfig]{article}
 \title{Analysis of $D_{s}^{+}{\rightarrow}{\phi}{\pi}^{+}$ beyond naive
factorization
 \thanks{Supported in part by National Natural Science Foundation of
        China and State Commission of Science and Technology of China}}
 \author{Haijun Gong  $^{b}$ \ \
        Junfeng Sun  $^{b}$  \ \
        Dongsheng Du $^{a,b}$ \\
 {\small\em a. CCAST (World Laboratory),
               P.O.Box 8730,
               Beijing 100080, China} \\
 {\small\em b. Institute of High Energy Physics, 
               Chinese Academy of Sciences,} \\ 
 {\small\em    P.O.Box 918(4), 
               Beijing 100039, China}
  \footnote{Email:gonghj@mail.ihep.ac.cn,
                  sunjf@mail.ihep.ac.cn,
                  duds@mail.ihep.ac.cn,}}
 \date{\today}
 
\begin{document}
 \maketitle
 \begin{abstract}
    We analyze the decay $D_s{\rightarrow}{\phi}{\pi}$ with QCD
factorization in the heavy quark limit. The nonfactorizable contributions,
including hard spectator contribution are discussed and numerical results
are presented. Our predictions on the branching ratio of the decay are in 
agreement with the experiment. We also use a pure phenomenological
method to estimate the branching ratio for $D_s{\rightarrow}{\phi}{\pi}$
with the existed $D^{0}{\rightarrow}K^{*}{\pi}$ data. 
 \end{abstract}

 {\bf PACS}:13.25.Ft 12.38.Bx

 \vfill

 \section{Introduction}
 \label{sec:introdution}
 Both CLEO $\cite{cleo}$ and BES $\cite{BES}$ have reported their direct
 model-independent measurements for the $D_{s}{\rightarrow}{\phi}{\pi}$
 branching fraction:  
 \begin{eqnarray}
 {\cal B}r(D_s\to\phi\pi)=\cases{
 (3.59\pm0.77\pm0.48)\times 10^{-2} & CLEO, \cr \nonumber
 (3.9^{ +5.1 + 1.8}_{ -1.9 - 1.1})\times 10^{-2} & BES.} \cr 
 \end{eqnarray}
 The average branching ratio of $D_{s}{\rightarrow}{\phi}{\pi}$ is
 $(3.6 \pm0.9 ){\times}10^{-2}\cite{PDG}$. \ 

  The precise estimation of the branching ratio for the
decay $D_{s}{\rightarrow}{\phi}{\pi}$ is very important. First, it is
difficult to measure the absolute branching ratio of $D_{s}{\rightarrow}{\phi}{\pi}$
because we do not know the fraction of $D_{s}^{+}D_{s}^{-}$ pairs 
production in $e^{+}e^{-}$ annihilation in comparison with
$D{\overline{D}}$ pairs ( BES used $e^{+}e^{-}\to D_{s}^{+}D_{s}^{-}$ to
obtain the first direct model-independent measurement of the
$D_{s}{\rightarrow}{\phi}{\pi}$ branching fraction, however, with only two "double-tagged" events ). But
we need to know the branching ratio for the study of B decays such as
$B{\rightarrow}D_{s}X$ etc. Moreover, most of the measurements of the
$D_s$ meson branching fractions are normalized to the clean
$D_{s}{\rightarrow}{\phi}{\pi}$ channel. Second, theoretically, the decay
of $D_{s}{\rightarrow}{\phi}{\pi}$ is dominated by spectator diagram with
external emission of pion. This is easier to handle compared with other
exclusive non-leptonic decay channels.

Previous calculations for the branching ratio ${\cal B}r(D_{s}
{\to}{\phi}{\pi})$ are based on the naive factorization approach which 
is proposed by Bauer, Stech and Wirbel(BSW) $\cite{BSW}$. But in BSW
approach, non-factorizable effects can not be calculated, they have to be
parameterized by an effective color number $N_c^{eff}$ which is treated
as a free parameter. Moreover, results obtained with BSW approach still
depend on renormalization scale and scheme. The authors in
$\cite{0012120}$ examine the $D_{s}{\rightarrow}{\phi}{\pi}$ amplitude
through a constituent quark-meson model. With this model, the
calculated decay width $\Gamma(D_{s}{\rightarrow}{\phi}{\pi})$ is larger 
than the experimental data. Paver and Riazuddin$\cite{0107330}$ studied 
$D_{s}{\rightarrow}{\phi}{\pi}$ in a valence quark triangle model, 
incorporating chiral symmetries, the result is compatible with the 
experimental data. In $\cite{9507358,9501246}$, the authors considered the
contribution from the color octet:
$\left\langle\phi\pi^+|H_w^{8}|D_s^+\right\rangle$, where, $ H_w^{(8)} 
\equiv {1 \over 2} \sum_{a}^{}{(\bar{u}\lambda^ac)(\bar{s}\lambda^ad)}$. 
But they all introduced some new parameters, so they brought new
theoretical uncertainties.

 In the past years, Beneke, Buchalla, Neubert and Sachrajda developed QCD
 factorization (QCDF) approach $\cite{BBNS}$ to calculate the hadronic matrix 
 elements of B decays in the heavy quark limit. It has been used for many B decays modes
 $\cite{BBNS,Du}$ with interesting results. In the present paper,
 we will follow this method to calculate the branching ratio for 
 $D_{s}{\rightarrow}{\phi}{\pi}$. In the heavy quark limit
$m_{c}{\gg}{\Lambda}_{QCD}$, non-factorizable contributions are 
considered from the first principle. In $D_{s}{\rightarrow}{\phi}{\pi}$
decay, the hadronic matrix elements can be represented as:
 \begin{eqnarray}
 {\langle}\pi\phi{\vert}Q_{i}({\mu}){\vert}D_s{\rangle}&=&
 {\langle}\pi{\vert}J_{1}{\vert}0{\rangle}  
 {\langle}\phi{\vert}J_{2}{\vert}D_s{\rangle}
 {\cdot}[1+{\sum}r_{n}{\alpha}_{s}^{n}+{\cal O}({\Lambda}_{QCD}/m_{c})].
 \label{eq:qcd-0}
 \end{eqnarray}
 The naive factorization corresponds to neglecting the ${\cal
O}({\alpha}_s)$ corrections and the power corrections in
${\Lambda}_{QCD}/m_{c}$. Although $m_c$ is not as large as $m_b$, we still
hope that the QCD factorization approach in the heavy quark limit can also
give a reasonable description of $D_s$ meson hadronic decays. With this
method, we analyze $D_{s}{\rightarrow}{\phi}{\pi}$ decay and compare the
results with those obtained with naive factorization. Finally we use
existed data of $D^{0}{\rightarrow}K^{*-}{\pi}^+$ to estimate the
branching ratio ${\cal B}r(D_{s}{\rightarrow}{\phi}{\pi})$ in a model
independent way. 
   \section{$D_{s}{\rightarrow}{\phi}{\pi}$ in QCD Factorization}
 \label{sec:framework}
 \label{sec:Hamiltonian}
 The low energy effective Hamiltonian for $D_{s}{\rightarrow}{\phi}{\pi}$ 
 can be expressed as follows :
 \begin{eqnarray}
 {\cal H}_{eff}&=&\frac{G_{F}}{\sqrt{2}}
 V_{cs}^{*}V_{ud} [ C_{1}({\mu}) Q_{1}({\mu})+C_{2}(\mu)Q_{2}({\mu}) ].
 \end{eqnarray}
The four-quark local operators $Q_{1,2}$ are
 \begin{equation}
 \begin{array}{ll}
 Q_{1}=({\bar{s}}_{\alpha}c_{\beta} )_{V-A}
            ({\bar{u}}_{\beta} d_{\alpha})_{V-A} \\
 Q_{2}=({\bar{s}}_{\alpha}c_{\alpha})_{V-A}  
            ({\bar{u}}_{\beta} d_{\beta} )_{V-A},\\
\end{array}
 \label{eq:operator}
 \end{equation}
 where ${\alpha}, {\beta}$ are the color indices of $SU(3)_{C}$. Wilson
coefficients $C_i(\mu)$ are  universal, process-independent and calculable
with the renormalizaion  group improved perturbative theory, their
${\mu}$-dependence are  expected  to be cancelled by the hadronic matrix
elements. The leading order (LO) and next-to-leading order (NLO)
corrections to $C_i(\mu)$ have been  presented in $\cite{Wilson}$. In the
naive dimensional regularization (NDR) scheme, we give the  numerical
values for $C_{i}({\mu})$ at three renormalization scales in  Tab.1. In
Fig.1, we also display the dependence of $C_i(i= 1,2)$ on  ${\mu}$ in the
LO and NLO approximation. We will take the values of $C_i(i= 1,2)$ at NLO 
for our forthcoming calculations.

\begin{figure}[t]
   \vspace{-3.05cm}
   \hspace*{1.0mm}
   \epsfysize=25cm
   \epsfxsize=15cm
   \centerline{\epsffile{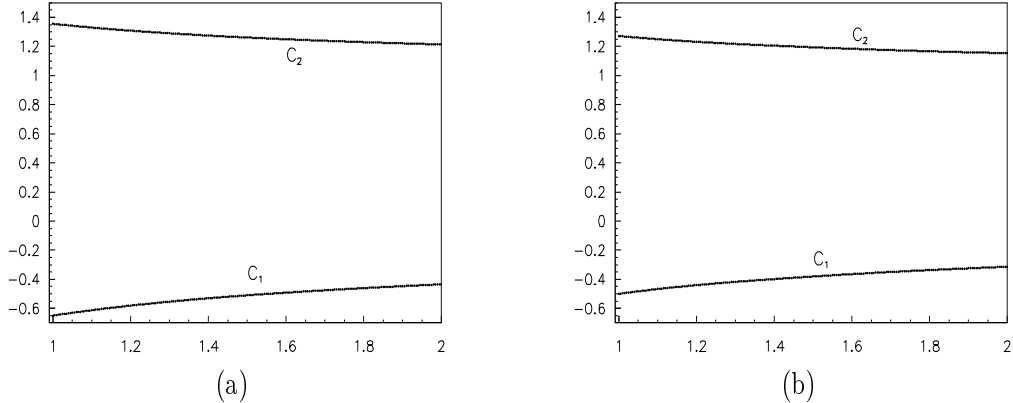}}
   \vspace*{-17.5cm}
\caption[dummy]{\small Dependence of the Wilson Coefficients $C_1,C_2$ on 
the renormalization scale $\mu$ at leading order (a) and next-to-leading
order (b). 
 \label{fig}}
\end{figure}

\begin{table}[tb]
 \label{table1} 
 \begin{center} 
 \caption{Wilson coefficients in NDR scheme. The input parameters 
 in numerical calculations are fixed: 
 ${\alpha}_{s}(m_{Z})= 0.1185$, 
 $ {\alpha}_{em}(m_{W})= 1/128$, 
  $m_{W}= 80.42$ GeV, 
  $m_{Z}= 91.188$ GeV, 
  $m_{t}= 168.2$ GeV, 
  $m_{c}= 1.45$ GeV . }
  \begin{tabular}{|l|c|c|c|c|} 
      \multicolumn{5}{c}{ }
      \\  \hline            
      \multicolumn{1}{|c|}{${\mu}$}              
    & \multicolumn{2}{|c|}{$C_1({\mu})$}                   
    & \multicolumn{2}{|c|}{$C_2({\mu})$}                   
       \\ \cline{2-5}                            
      \multicolumn{1}{|c|}{}                     
    & \multicolumn{1}{|c|}{LO}                   
    & \multicolumn{1}{|c|}{NLO}                  
    & \multicolumn{1}{|c|}{LO}                   
    & \multicolumn{1}{|c|}{NLO} 
    \\ \hline                                  
    ${\mu}= 1$GeV                                 
     &-0.650  &-0.500                              
     &1.356   &1.272  \\                          
    ${\mu}=m_c$                                 
      &-0.520  &-0.390                 
      &1.268  &1.200  \\                         
    ${\mu}=2$GeV                                
      &-0.435  &-0.314                             
      &1.214   &1.153  \\    
       \hline                  
       \end{tabular}                             
       \end{center}                              
       \label{wilson}                            
       \end{table}                               
The decay constant and form factors are defined by $\cite{BSW}$ :
\begin{eqnarray}
\langle\pi^+|(\bar{u}d)_{(V-A)}^{\mu}|0\rangle&=&-i f_{\pi}
p_{\pi}^\mu,\\
\langle\phi|(\bar{s}c)_{{\mu}(V-A)}|D_s^+\rangle&=&-i \left\{
(m_{D_s} + m_\phi)
\varepsilon_\mu^* A_1^{D_s\phi}
(q^2) - {\varepsilon^*.q \over m_{D_s} + m_\phi} (p_{D_s} + p_\phi)_\mu
A_2^{D_s\phi}(q^2) \right.
\nonumber \\
&-& \left.2 m_\phi {\varepsilon^*.q \over q^2}  q_\mu A_3^{D_s\phi}(q^2)+
{\varepsilon^*.q \over q^2}
(2m_\phi)q_\mu A_0^{D_s\phi}(q^2) \right\}  \nonumber \\
&+& {2 \over m_{D_s} + m_\phi}\varepsilon_{\mu \nu \rho
\sigma}\varepsilon^{*\nu}p^\rho_{D_s} p^\sigma_\phi V^{D_s\phi}(q^2),
\label{eq:SL}
\end{eqnarray}
where $q_\mu = (p_{D_s} - p_\phi)_\mu$ and
\begin{eqnarray}
A_0^{D_s\phi}(0)&=&A_3^{D_s\phi}(0), \\
2 m_\phi A_3^{D_s\phi}(0)&=&(m_{D_s} + m_\phi) A_1^{D_s\phi}(0) - (m_{D_s}
- m_\phi) A_2^{D_s\phi}(0).
 \label{eq:A2}
\end{eqnarray}
The relations (6) - (\ref{eq:A2}) ensure that there is no
kinematical singularity in the matrix element at $q^2= 0$.\

  Under naive factorization, using Eq. (4) - (\ref{eq:A2}),
the decay amplitude of $D_{s}{\rightarrow}{\phi}{\pi}$ reads \\
 \begin{eqnarray}
{\cal
A}(D_{s}{\rightarrow}{\phi}{\pi})={\sqrt{2}}G_{F}V_{cs}^{*}V_{ud}f_{\pi}m_{\phi}A_{0}^{D_s{\phi}}
(m_{\pi}^2)({\epsilon}^{*}\cdot p_{D_s})\cdot a_{2},
 \label{eq:A}
 \end{eqnarray}
 where $a_{2}=C_{2}+\frac{1}{N_{c}^{eff}}C_{1}$, $N_{c}^{eff}$ is the
number of colors. The form factor $A_{0}^{D_s{\phi}}$ is defined by
(\ref{eq:SL}). From Eq. (\ref{eq:A}) we can see that the amplitude 
depends on the renormalization scale ${\mu}$, because the Wilson
coefficients $C_1({\mu})$, $C_2({\mu})$, and hence $a_1$, $a_2$ depend on
${\mu}$, whereas the decay constant and form factor are independent of
${\mu}$. So the amplitude ${\cal A}(D_{s}{\rightarrow}{\phi}{\pi}$) 
is $\mu$-dependent. On the other hand, it does not consider the
nonfactorizable effects. If we calculate it with QCD factorization, take
all the high order corrections into account, $a_i$ and the amplitude
${\cal A}(D_{s}{\rightarrow}{\phi}{\pi}$) will be ${\mu}$ independent. 
In our paper, we calculate it only to the order of ${\alpha}_s$, so $a_i$
and the amplitude ${\cal A}(D_{s}{\rightarrow}{\phi}{\pi}$) still depend
on $\mu$, but the dependence is less sensitive to $\mu$. With these
preliminaries, we now analyze $D_{s}{\rightarrow}{\phi}{\pi}$ with QCD
factorization.

 In $D_{s}{\rightarrow}{\phi}{\pi}$ decay, the emitted meson ${\pi}$ is light,
 the hadronic matrix elements can be written as:
 \begin{eqnarray}
 {\langle}\pi\phi{\vert}Q_{i}({\mu}){\vert}D_s{\rangle}&=&
 A^{D_s\phi}_0(0){\int}_{0}^{1}dx\;T^{I}_{i}(x){\Phi}_{\pi}(x)
 \nonumber \\ & &
 +{\int}_{0}^{1}d{\xi}\;dx\;dy\;T_{i}^{II}({\xi},x,y)
  {\Phi}_{D_s}({\xi}){\Phi}_{\pi}(x){\Phi}_{\phi}(y).
 \label{eq:qcd}
 \end{eqnarray}
 $A^{D_s\phi}_0(0)$ denotes the $D_s{\to}\phi$ transition form factor,
 ${\Phi}_D({\xi})$, ${\Phi}_{\pi}(x)$ and ${\Phi}_{\phi}(x)$ label 
light-cone distribution amplitudes (LCDAs) of $D_s$, ${\pi}$ and ${\phi}$
meson respectively. $T^{I,II}_{i}$ denote hard-scattering kernels which are
calculable in perturbative theory. Neglecting the ${\cal
O}({\Lambda}_{QCD}/m_{c})$  corrections, $T^{I,II}_{i}$ are hard gluon
exchange dominant. Other  non-perturbative contributions are contained in
the LCDAs of mesons or the form factor. The second term in Eq.
(\ref{eq:qcd}) represents the hard spectator contribution. \

  We next proceed to calculate the nonfactorizable effects in the
 $D_{s}^{+}{\rightarrow}{\phi}{\pi}^{+}$ with QCDF approach.
Then in heavy quark limit, for simplicity, we will neglect the masses of light
quarks and ${\pi}$. We consider the vertex corrections and hard spectator
interactions depicted in Fig.2. The technique is similiar to that of the
$B{\rightarrow}{\pi}{\pi}/K$ mode, readers can be referred to
$\cite{BBNS}$ for details. As in $\cite{BBNS}$, we obtain the QCD
coefficients $a_{i}$(i= 1, 2 ) at NLO in NDR scheme. Then the coefficients
$a_i$ are given as 
\begin{eqnarray}
a_1 &=& C_1 +\frac{C_2}{N} +\frac{\alpha_s}{4\pi} \frac{C_F}{N} C_2
F, \nonumber \\
a_2 &=& C_2 +\frac{C_1}{N} +\frac{\alpha_s}{4\pi} \frac{C_F}{N} C_1
F.
\end{eqnarray}
Here $N= 3$ $(f= 4)$ is the number of colors (flavors), and
$C_F= \frac{N^2-1}{2N}$ is the factor of color. We define the symbols in
the above expressions as the same as Beneke's, which are 
\begin{eqnarray}
 F &=& -18 -12\ln \frac{\mu}{m_c} + f_{I} + f_{II}, \\
 f_I&=&\int^1_0 \!\!\!dx\,g(x)\Phi_\pi(x),\quad
\end{eqnarray}
with the hard-scattering function
\begin{eqnarray}
\label{g8x}
g(x)= 3\frac{1-2x}{1-x}\ln x-3 i\pi.\nonumber
\label{ggsx}
\end{eqnarray}

The hard spectator scattering contribution is given by
\begin{equation}\label{hbpi}
f_{II}= \frac{4\pi^2}{N}\frac{f_\phi f_{D_s}}{A_0^{D_s\phi}(0)m^2_{D_s}}
\int^1_0\!\!\!d\xi\,\frac{\Phi_{D_s}(\xi)}{\xi}
\,\int^1_0\!\!\!dx\,\frac{\Phi_\pi(x)}{x}
  \int^1_0\!\!\!dy\,\frac{\Phi_\phi(y)}{y} ,
\end{equation}
where $f_\phi$ ($f_{D_s}$) is the $\phi$ ($D_s$) meson decay constant,
$m_{D_s}$ the mass of $D_s$ meson, $A_0^{D_s\phi}(0)$ the $D_s\to\phi$
transition form factor at zero momentum transfer, and $\xi$ the light-cone
momentum fraction of the spectator quark in the $D_s$ meson, $f_{II}$
depends on the wave function $\Phi_{D_s}$ through the integral 
\begin{equation}
\int^1_0d\xi \,\Phi_{D_s}(\xi)/\xi
= m_{D_s}/\lambda_D.
\end{equation}
 This introduces a new hadronic parameter $\lambda_D$, $\lambda_D$ is 
of order $\Lambda_{QCD}$, we take $\lambda_D= 335$ MeV here. 

\begin{figure}[t]
   \vspace{-3.05cm}
   \hspace*{1.0mm}
   \epsfysize=20cm
   \epsfxsize=15cm
   \centerline{\epsffile{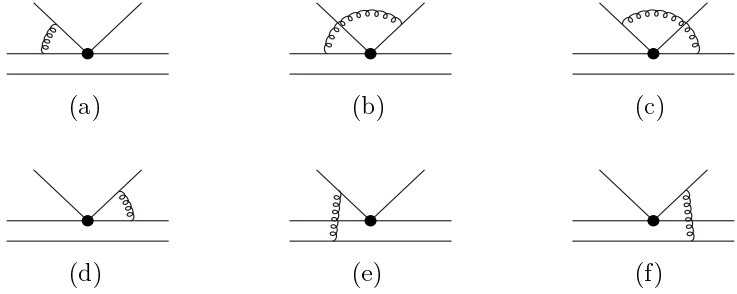}}
   \vspace*{-13.7cm}
\caption[dummy]{\small Order of $\alpha_s$ corrections to the hard
scattering kernels $T^I_i$ and $T^{II}_i$. The two lines directed
upwards represent the two quarks that make up $\pi$. These diagrams are 
called vertex corrections for Fig.(a)-(d) and hard spectator diagrams for
Fig.(e)-(f) respectively.
 \label{fig2}}
\end{figure}

From the expression (10) of the QCD coefficients $a_i(i= 1, 2)$, with
the renormalization group equation for Wilson coefficients $C_i(\mu)$ at 
leading order logarithm approximation $\cite{Wilson}$ : 
 \begin{equation}
{d\,C_i(\mu)\over d\ln \mu }=\,{\alpha_s\over
4\pi}\,\gamma^T_{ij}C_i(\mu),
 \end{equation}
where $\gamma$ is the anomalous dimension matrix, we find $\frac{{\bf
d}a_{i}}{{\bf d}{\ln}{\mu}}= 0$ ($i= 1,2$) at the order of ${\alpha}_{s}$, 
this makes the $\mu$-dependence of the decay amplitude calculated with 
QCDF approach less sensitive than that calculated with naive
factorization. This point can also be seen roughly from the data in Tab.2
and Fig.3 - 4. But there are still uncertainties in the calculation, such
as the form of wave functions and unknown form factor $A_0^{D_s\phi}$. 

Notice that in the decay $D_s\rightarrow \phi\pi$, using the isospin
analyses $\cite{sinha,rosner}$, we find that the final state involves
only a single isospin, so there is no interference effects from the final
state interactions (FSI) when we calculate the branching ratio of
$D_s\rightarrow \phi\pi$.

In the $D_s$ rest frame, the two body decay width is
\begin{equation}  
\Gamma(D_s\rightarrow \phi\pi)=\frac {1}{8\pi}
\vert A(D_s\rightarrow \phi\pi) \vert ^2
\frac{\vert p \vert}{m_{D_s}^2},
\end{equation}
where
\[ {\vert}p{\vert}=\frac{\sqrt{\Big[m_{D_s}^{2}-(m_{\phi}+m_{\pi})^{2}
   \big]\Big[m_{D_s}^{2}-(m_{\phi}-m_{\pi})^{2}\Big]}}{2m_{D_s}} 
\]
     is the magnitude of the momentum of $\phi$ meson. With the
approximation $m_\pi^2/m_{D_s}^2 \approx 0$, the decay width is given by
\begin{equation}
\Gamma\left( D_s^+ \rightarrow \phi \pi^+ \right) = {G_F^2 m_{D_s}^5\over 32
\pi} |V_{cs}|^2
|V_{ud}|^2 |a_2|^2 \left( {f_{\pi} \over m_{D_s}} \right)^2\left(1-\left(
{m_{\phi} \over m_{D_s}}\right)^2 \right)^3(A_0^{D_s\phi}(0))^2.
\end{equation}

The corresponding branching ratio is given by 
\begin{equation}
Br(D_s\rightarrow \phi\pi)
=\frac{\Gamma(D_s\rightarrow \phi\pi)}{\Gamma_{total}},\
 \Gamma_{total}=\frac{1}{\tau_{D_s}}.
 \end{equation} 
In our numerical calculations, we will take the
following values for the relevant input parameters $\cite{PDG}$: 
$V_{cs}$ = $V_{ud}$ = 0.975, $f_{\pi}$ = 131 MeV, $f_{\phi}$ = 233 MeV,
$m_c$ = 1.45 GeV, $f_{D_s}$= 280 ${\pm}$ 19 ${\pm}$ 28 ${\pm}$ 34 MeV. As
for the form factor $A_0^{D_s\phi}(0)$, for lack of experimental data, we
use the value taken from the Ref. $\cite{BSW}$\ $A_0^{D_s\phi}$(0) = 0.70.
For mass of the mesons, we use $m_{D_s}$= 1968.6 ${\pm}$ 0.6 MeV,
$m_{\phi}$= 1019.417 ${\pm}$ 0.014 MeV. If not stated otherwise, we shall
use the central values as the default values in our later calculations.

For distribution amplitude of ${\pi}$, two kinds of the wave functions are
used, one is the asymptonic form $\cite{BBNS}$ $\Phi_{\pi}(x) = 6x(1-x)$,
the other is delta-function ${\Phi}_{\pi}(x) = {\delta}(x-\frac{1}{2})$. 
In Tab.2 we list the values of $a_1, a_2$ and branching ratio(${\cal B}r$)
at ${\mu}$= 1 GeV, $m_c$, and 2 GeV with different wave functions of
$\pi$. The numerical results which are calculated with BSW approach (
Where we take $N^{eff}_c = {\infty}$ because the experimental data of 
MARK III for charm decays do not show color suppression $\cite{hauser}$ )
are also listed for comparison. 
\begin{table}[htb]  
\begin{center}
  \caption{The values of $a_i$ and ${\cal B}r$ at ${\mu}$= 1 GeV, $m_c$
and 2 GeV calculated with QCDF and BSW approach ($N^{eff} = {\infty}$ ).
For QCDF, we calculate the spectator contribution with three forms:
$f_{II}$ = 0 and two different wave functions of ${\pi}$. In the QCDF
columms, the values in the parentheses are those with $\Phi_{\pi}(x) =
6x(1-x)$, the values in the brackets are those with ${\Phi}_{\pi}(x) =
{\delta}(x-\frac{1}{2})$. } 
\vspace{7mm} 
\begin{tabular}{|l|c|c|c|c|c|c|} \hline            
      \multicolumn{1}{|c|}{}              
    & \multicolumn{2}{|c|}{$a_1$}                   
    & \multicolumn{2}{|c|}{$a_2$} 
     & \multicolumn{2}{|c|}{${\cal B}r \ \%$}                 
       \\ \cline{2-7}                            
      \multicolumn{1}{|c|}{${\mu}$ }                     
    & \multicolumn{1}{|c|}{BSW}                   
    & \multicolumn{1}{|c|}{QCDF}                  
    & \multicolumn{1}{|c|}{BSW}                   
    & \multicolumn{1}{|c|}{QCDF}
    & \multicolumn{1}{|c|}{BSW}                   
    & \multicolumn{1}{|c|}{QCDF}
    \\ \hline                                  
     &        & -0.396 - 0.215 $i$ 
     &        & 1.231 + 0.084 $i$ 
     &        & 3.53    \\ 
    ${\mu}$= 1 GeV
     &-0.500  &( -0.071 - 0.215 $i$ )                               
     &1.272   &( 1.103 + 0.084 $i$ )
     &3.75    &( 2.84 )   \\                          
     &        &[ -0.168- 0.215 $i$ ]
     &        &[ 1.141 + 0.084 $i$ ]
     &        &[ 3.03 ]   \\  \hline     
      &        & -0.284 - 0.150 $i$ 
      &        & 1.165 + 0.049 $i$ 
      &        & 3.15   \\ 
    ${\mu}= m_c$
      &-0.390  &( -0.058 - 0.150 $i$ )                 
      &1.200  &( 1.092 + 0.049 $i$ )  
      &3.33   &( 2.77 ) \\                         
      &        &[ -0.125 - 0.150 $i$ ]
      &        &[ 1.112 + 0.049 $i$ ]
      &        &[ 2.88 ]  \\ \hline
      &       & -0.213 - 0.119 $i$ 
      &       & 1.126 + 0.033 $i$ 
      &       & 2.94 \\
   ${\mu}$= 2 GeV 
      &-0.314  &( -0.032- 0.119 $i$ )                             
      &1.153   &( 1.077 + 0.033 $i$ ) 
      &3.08    &( 2.69 )\\ 
      &       &[ -0.086 - 0.119 $i$ ]
      &       &[ 1.091 + 0.033 $i$ ]
      &       &[ 2.76 ]\\  \hline   
       \end{tabular}                             
      \label{table1}
       \vspace{5mm} 
       \end{center} 
       \end{table}

It is necessary to note that the QCDF approach gives $a_i(i= 1,2)$ an 
imaginary part, which comes from the gluon exchange of the quarks $u$
and ${\overline d}$ in ${\pi}$ with the $s$ quark in $\phi$ (see Fig.2
(c)-(d)). Moreover, the imaginary parts of $a_i(i= 1, 2)$ have no
relations with $f_{II}$. From the numerical values summarized in Tab.2, we
find that the vertex correction in Fig.2 (a)-(d) is about 5$\sim7\%$, the
hard-spectator diagrams can reduce over 10$\%$ of the values obtained with 
BSW approach. And the coefficients $a_i(i = 1, 2)$ are less sensitive to
the choice of the wave functions. In Fig.3, we depict the dependence of
$a_1, a_2$ and ${\cal B}r$ on scale ${\mu}$ when considering the vertex
corrections (but neglecting the hard spectator contribution), we also show
the results calculated by BSW approach for comparison. The horizontal
solid lines in Fig.3(b) show the experimental branching ratio at
1${\sigma}$ level. It is clear that the scale dependence of the values 
calculated with QCDF approach are milder than that with BSW approach. But
the $\mu$ dependence still exists, the reason is that we calculate 
$a_i$ only at one-loop level, the source of $\mu$ dependence is from the
high order effects. When considering the contributions from the high order
corrections in $\alpha_s$ or $\Lambda_{QCD}/m_c$, the $\mu$ dependence of
our predictions will be further reduced.

\begin{figure}[t]
   \vspace{-3.05cm}
   \hspace*{1.0mm}
   \epsfysize=25cm
   \epsfxsize=15cm
   \centerline{\epsffile{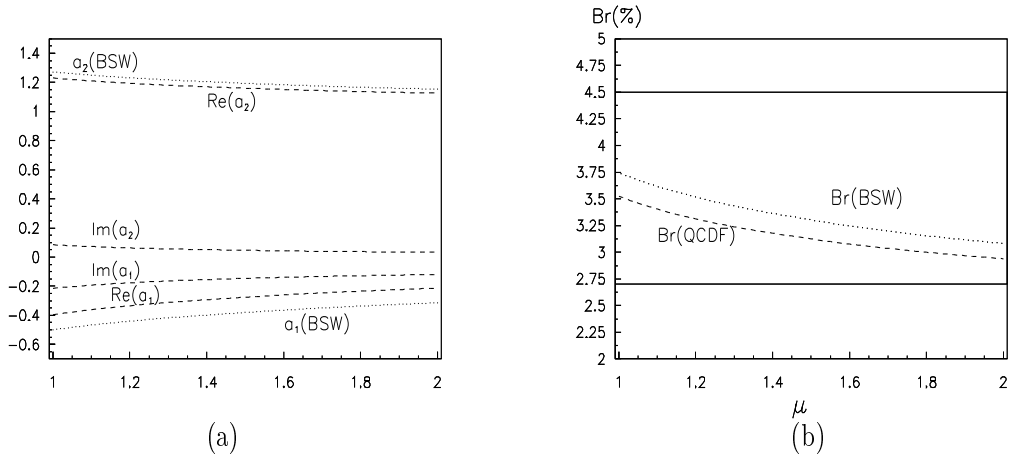}}
   \vspace*{-17.2cm}
\caption[dummy]{\small Dependence of $a_1, a_2$ and ${\cal B}r$ on the
renormalization scale $\mu$ in BSW and QCDF($f_{II}$ = 0). The dotted
and dashed lines correspond to the values obtained with BSW and QCDF
approach.
 \label{fig3}}
\end{figure}

In Fig.4, we compare the results which are calculated with different wave
functions of $\pi$ when considering the hard-spectator contribution in
QCDF. It shows again that $a_1, a_2$ and ${\cal B}r$ are less sensitive to
the selection of the wave function of ${\pi}$, moreover, their
$\mu$-dependence is furthur reduced.  From Fig.3 and Fig.4, we find that
the results obtained with QCD factorization approach fall in the 1${\sigma}$ 
allowed region from the central experimental value 3.6${\times}10^{-2}$,
regardless of the seletion of the function of ${\pi}$. Though the
branching ratios with BSW approach are also within the 1${\sigma}$ region,
this approach takes $N^{eff}_c = \infty$ in order to fit the experimental
data, so it is more phenomenological in comparison with QCDF approach.
From Fig.3 and Fig.4, we can see apparently that our predictions with
QCDF approach are small compared with the values obtained with BSW
approach.
\begin{figure}[htb]
   \vspace{-1.05cm}
   \hspace*{1.0mm}
   \epsfysize=25cm
   \epsfxsize=15cm
   \centerline{\epsffile{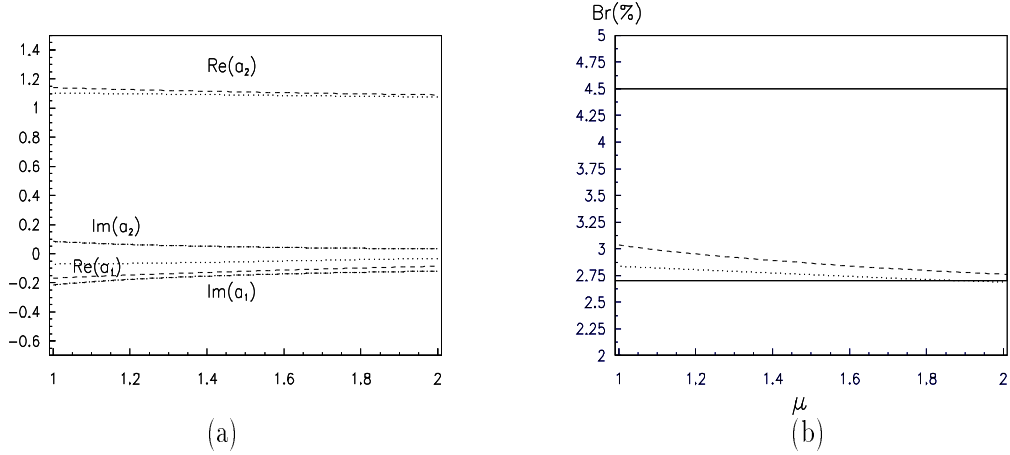}}
   \vspace*{-17.2cm}
\caption[dummy]{\small Dependence of $a_1, a_2$ and ${\cal B}r$ on the
renormalization scale $\mu$ in QCDF with different function of $\pi$.
The dotted and dashed lines correspond to the values obtained with 
$\Phi_{\pi}(x) = 6x(1-x)$ and ${\Phi}_{\pi}(x) = {\delta}(x-\frac{1}{2})$.   

 \label{fig4}}
\end{figure}
 \section{Direct estimation of ${\cal B}r(D_{s}{\rightarrow}{\phi}{\pi})$ }
\label{sec:direct}
Now we estimate the ${\cal B}r(D_{s}{\rightarrow}{\phi}{\pi})$ directly
from the existed data of $D^{0}{\rightarrow}K^{*-}{\pi}^{+}$. Assuming
spectator diagram dominance, $D_{s}^{+}{\rightarrow}{\phi}{\pi}^{+}$
can go through quark decay diagram depicted in Fig.5(a), the decay width
is (16). Using the experimental data listed in Sec.2, we get
${\vert}p{\vert}$= 0.720.

Consider the decay $D^{0}{\rightarrow}K^{*-}{\pi}$ which proceeds
dominantly through diagram Fig.5(b), obviously, in Fig.5, diagram (a) and
(b) are very similiar, if ${\overline s}$ in (a) is replaced by
${\overline u}$, we will get (b). In addition, the particle decay width of
$D^{0}{\rightarrow}{K^*}{\pi}$ is 
\begin{equation}  
\Gamma(D^0\rightarrow K^*\pi)=\frac {1}{8\pi}
\vert A(D^0\rightarrow K^*\pi) \vert ^2
\frac{\vert p{\prime} \vert}{m_{D^0}^2},
\end{equation}
where ${\vert}p{\prime}{\vert}$ = 0.719. The momentum of $K^{*}$ in
the $D^0$ rest frame is almost the same as that of $\phi$ in 
$D_{s}^{+}{\rightarrow}{\phi}{\pi}^{+}$. So the Lorentz contraction
effects of the wave function of ${\phi}$ and $K^{*}$ are nearly the same.
We know that the decay amplitudes $ A(D_s\rightarrow \phi\pi)$ and $
A(D^0\rightarrow K^{*}\pi)$ are proportional to the wave function overlap
integrates of $D_{s}^{+}-{\phi}$ and $D^0-K^{*}$ respectively. Moreover, 
${\vert}p{\vert}$= 0.720 and ${\vert}p{\prime}{\vert}$ = 0.719 mean that
these overlap integrates are almost the same under the condition of SU(3)
symmetry. The SU(3) symmetry breaking effects in the case of
$D_{s}^{+}{\rightarrow}{\phi}{\pi}^{+}$ and $D^0\rightarrow K^{*-}\pi^{+}$
should be fairly small. So in approximation, we can have  
\begin{equation}  
\vert A(D_s\rightarrow \phi\pi) \vert {\approx} \vert A(D^0\rightarrow
K^{*-}\pi^{+}) \vert. 
\end{equation}
\begin{figure}
\vspace{-2.05cm}
 \hspace*{-1mm}
\epsfysize=20cm
 \epsfxsize=18cm
 \centerline{\epsffile{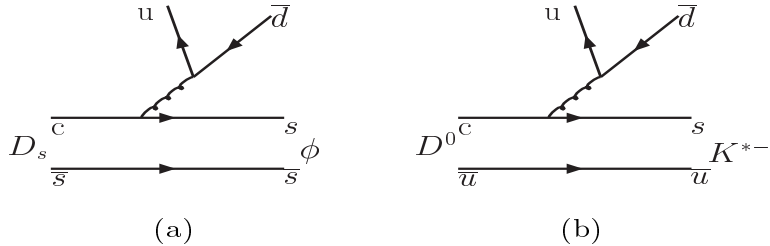}}
\vspace*{-15cm}
\caption[dummy]{\small Diagrams for the decay
$D_{s}{\rightarrow}{\phi}{\pi}$ and $D^0\rightarrow K^*\pi$.
 \label{fig3}}
 \end{figure}

Using the experimental data $\cite{PDG}$:
 \begin{eqnarray}
 {\tau}(D^0) = (0.4126\pm0.0028){\times} 10^{-12}s,\nonumber \\
 {\cal B}r(D^0\rightarrow K^{*-}\pi^+)= (5.0\pm0.4)\%, \nonumber \\
 \tau (D_s)= (0.496 ^{+0.010} _{-0.009}) \times 10^{-12}s, \nonumber 
  \end{eqnarray}
with Eq.(18) - (20), we obtain ${\cal B}r(D_s\to \phi\pi)\approx$ 
(5.40 ${\pm}$ 0.45)$\%$, where the error comes from that of the data of
$\tau(D^0)$, ${\cal B}r(D^0\rightarrow K^{*-}\pi^+)$ and $\tau(D_s)$.
It is a little outside the one $\sigma$ allowed region from the central
experimental value 3.6${\times} 10^{-2}$.
 \section{Conclusions}
    We have analyzed the decay $D_s{\rightarrow}{\phi}{\pi}$ with QCD
factorization in the heavy quark limit. We calculate the nonfactorizable
contributions, including vertex correction, hard-spectator contribution. 
These nonfactorizable contributions can give over $10\%$ corrections to
naive factorization. Moreover, according to our calculations, the branching 
ratios with QCDF approach is not sensitive to the choice of the wave
function of pion. Our predictios are in agreement with the present 
experimental data. The direct estimation of ${\cal B}r(
D_s{\rightarrow}{\phi}{\pi})$ from $D^{0}{\rightarrow}K^{*}{\pi}$ data
gives a bit larger result comparing with the present data. But the
measured data on ${\cal B}r(D_s{\rightarrow}{\phi}{\pi})$ are still rough, 
we need more data for drawing our final conclusion.
 \section*{Acknowledgemanets}
 We thank Dr. Deshan Yang and Guohuai Zhu for helpful discussions about 
QCD factorization. This work is Supported in part by National Natural
Science Foundation of China and State Commission of Science and 
Technology of China.
 
 \end{document}